\documentclass[12pt]{article}
\usepackage{bbm,latexsym}
\usepackage{amssymb,amsmath}
\usepackage{graphicx}
\usepackage{slashed}
\newcommand{\mc}{\mathcal{M}}
\allowdisplaybreaks
\begin{document}
\renewcommand{\thefootnote}{\fnsymbol{footnote}}

\title{
\normalsize \hfill UWThPh-2014-16 \\[10mm]
\LARGE Mass renormalization in a toy model \\
with spontaneously broken symmetry}

\author{
W.~Grimus\thanks{E-mail: walter.grimus@univie.ac.at},\;
P.O.~Ludl\thanks{E-mail: patrick.ludl@univie.ac.at}\;
and L. Nogu\'es\thanks{E-mail: esplebizonear@gmail.com}
\\[5mm]
\small University of Vienna, Faculty of Physics \\
\small Boltzmanngasse 5, A--1090 Vienna, Austria
}

\date{April 24, 2017}

\maketitle

\begin{abstract}
We discuss renormalization in a toy model with one fermion field and one
real scalar field $\varphi$, featuring a spontaneously broken discrete
symmetry which forbids a fermion mass term and a $\varphi^3$ term in the
Lagrangian. 
We employ a renormalization scheme which uses the $\overline{\mbox{MS}}$
scheme for the Yukawa and quartic scalar couplings and 
renormalizes the vacuum expectation value of $\varphi$
by requiring that the one-point function of the shifted field is zero.
In this scheme, the
tadpole contributions to the fermion and scalar selfenergies are canceled
by choice of the renormalization parameter $\delta v$ of the vacuum expectation
value. However, $\delta v$ and, therefore, the tadpole contributions reenter
the scheme via the mass renormalization of the scalar, in which place  
they are indispensable for obtaining finiteness.  
We emphasize that the above renormalization scheme 
provides a clear formulation of the hierarchy problem and allows a
straightforward generalization to an arbitrary number of fermion and scalar
fields. 
\end{abstract}

\newpage

\renewcommand{\thefootnote}{\arabic{footnote}}

\section{Introduction}

Our toy model is described by the Lagrangian
\begin{equation}\label{L}
\mathcal{L} = i \bar \chi_L \gamma^\mu \partial_\mu \chi_L +
\left( \frac{1}{2}\,y\, \chi_L^T C^{-1} \chi_L \varphi + 
\mbox{H.c.} \right) +
\frac{1}{2} \left( \partial_\mu \varphi \right) 
\left( \partial^\mu \varphi \right) - V(\varphi)
\end{equation}
with the scalar potential 
\begin{equation}
V(\varphi) = 
\frac{1}{2} \mu^2 \varphi^2 + \frac{1}{4} \lambda \varphi^4.
\end{equation}
It contains a real scalar field $\varphi$ and a Majorana fermion field
$\chi_L$. The symmetry
\begin{equation}\label{S}
\mathcal{S}: \quad \varphi \to -\varphi, \quad \chi_L \to i \chi_L
\end{equation}
forbids a tree-level mass term of the Majorana fermion and the term
$\varphi^3$ in the scalar potential. A possible phase in the Yukawa
coupling constant $y$ can be absorbed into the field
$\chi_L$. Therefore, without loss of generality we assume that $y$ is
real and positive.

Alternatively, we may consider a Dirac fermion in the toy model. In
this case, there are two different chiral fields $\chi_L$ and $\chi_R$
and the Lagrangian reads
\begin{equation}\label{LD}
\mathcal{L} = 
i \bar \chi_L \gamma^\mu \partial_\mu \chi_L +
i \bar \chi_R \gamma^\mu \partial_\mu \chi_R -
\left( y\, \bar\chi_L \chi_R \varphi + 
\mbox{H.c.} \right) +
\frac{1}{2} \left( \partial_\mu \varphi \right) 
\left( \partial^\mu \varphi \right) - V(\varphi).
\end{equation}
The transformation of the fermion field in equation~(\ref{S}) can 
for instance be modified to 
$\chi_L \to -\chi_L$ and $\chi_R \to \chi_R$, 
in order to forbid a fermion mass at the tree level.

We will assume $\mu^2 < 0$, which leads to spontaneous symmetry breaking of 
$\mathcal{S}$ with a
tree-level vacuum expectation value (VEV) of $\varphi$ given by
\begin{equation}\label{v}
v = \sqrt{-\frac{\mu^2}{\lambda}}.
\end{equation}
The tree-level masses
\begin{equation}\label{tm}
m_0 = yv
\quad \mbox{and} \quad 
M_0^2 = 2 \lambda v^2
\end{equation}
of the fermion and the scalar, respectively, ensue.

We are interested in the mass renormalization of this most simple
spontaneously broken toy model because we want to employ a special
renormalization scheme which imposes renormalization conditions on the
VEV~\cite{fleischer} and on the two coupling 
constants $y$ and $\lambda$, but not on the masses;\footnote{In 
  spontaneously broken theories it is natural to consider such a scheme. In
  effect, the very same philosophy is applied when the dependence of fermion
  masses on the renormalization scale is obtained from the 
  renormalization group running of the Yukawa couplings---see for
  instance~\cite{grimus}.} 
we want to discuss
this scheme in the most simple environment where we do not have to face the 
complications by gauge theories and propagator mixing.
We will explicitly compute the one-loop corrections to
equation~(\ref{tm}). Our motivation for considering this kind of mass
renormalization is the following:
\begin{enumerate}
\renewcommand{\labelenumi}{\roman{enumi}.}
\item
Due to the renormalization of the vacuum expectation value and the coupling
constants, the fermion and scalar masses must be automatically
finite. It is the purpose of these notes to work out how the
cancellation of divergences in the masses occurs. In particular, we
want to elucidate the role of tadpoles~\cite{fleischer,weinberg} in the context
of the scalar mass.
\item
At a later time, having in mind renormalizable 
flavour models for fermion masses and
mixing, we intend to extend the model to an
arbitrary number of fermion and scalar fields. In this case, there
are, in general, more Yukawa couplings than fermion masses and it
seems practical to renormalize coupling constants and VEVs instead of
masses. A further motivation for such a scheme is given by flavour symmetries
which may impose tree-level relations among the VEVs and among the Yukawa
couplings\footnote{Flavour symmetries have for instance the effect of
  enforcing a VEV or a Yukawa coupling constant to vanish or making two 
  VEVs or two Yukawa coupling constants equal at the tree level.}
and, therefore, also on the masses; such relations will obtain
finite corrections at the one-loop level.
\end{enumerate}

The paper is organized as follows. In section~\ref{renormalization} 
we describe the renormalization of the toy model. The one-loop fermion
and scalar selfenergies, together with the cancellation of infinities and
the corrections to equation~(\ref{tm}), are discussed in
sections~\ref{fermion} and~\ref{scalar}, respectively. Since for
simplicity we use 
the $\overline{\mbox{MS}}$ scheme~\cite{buras} 
for the renormalization of both the Yukawa and $\varphi^4$-coupling
constant, we have a dependence of renormalized quantities
on the renormalization scale parameter denoted by $\mc$ in our
paper, which is worked out in section~\ref{independence}.
The conclusions are presented in section~\ref{concl}.

\section{Renormalization}
\label{renormalization}
\paragraph{Dimensional regularization and renormalization:}
In the following, bare quantities carry a subscript $B$. 
Our starting point is the Lagrangian of equation~(\ref{L}), written in bare
fields, the bare coupling constants $y_B$, $\lambda_B$ and the
bare mass parameter $\mu^2_B$. We confine ourselves to the Majorana case
because the treatment of the Dirac case is completely analogous. 
Renormalization splits the bare Lagrangian into
\begin{equation}
\mathcal{L}_B = \mathcal{L} + \delta \mathcal{L},
\end{equation}
where $\mathcal{L}$ is the renormalized Lagrangian,
written in terms of renormalized quantities, and
$\delta \mathcal{L}$ contains the counterterms.
The renormalized fields $\chi_L$ and $\varphi$ are given by
the relations
\begin{equation}
\chi_{LB} = \sqrt{Z_\chi}\, \chi_L,
\quad
\varphi_B = \sqrt{Z_\varphi}\, \varphi.
\end{equation}
Using dimensional regularization, we are working in 
$d = 4 - \varepsilon$ dimensions. 
As usual, in order to have dimensionless coupling constants $y$ and $\lambda$
in $d$ dimensions, the coupling constants are rescaled
by
\begin{equation}\label{scale}
y \to y\, \mc^{\varepsilon/2}, \quad
\lambda \to \lambda\, \mc^{\varepsilon}
\end{equation}
with an arbitrary mass parameter $\mc$.
Therefore, the renormalized Lagrangian 
is identical with that of equation~(\ref{L}), provided that we make the
replacement of equation~(\ref{scale}). 
Now it is straightforward to write down 
the counterterms, subsumed in $\delta \mathcal{L}$:
\begin{eqnarray}
\nonumber
\delta \mathcal{L} &=& 
\delta_\chi\, i \bar \chi_L \gamma^\mu \partial_\mu \chi_L +
\delta_\varphi \frac{1}{2} \left( \partial_\mu \varphi \right) 
\left( \partial^\mu \varphi \right) 
\\ && +
\left( \frac{1}{2}\, \delta y \mc^{\varepsilon/2}\, 
\chi_L^T C^{-1} \chi_L + \mbox{H.c.} \right) 
\varphi - 
\frac{1}{2} \delta \mu^2 \varphi^2 - \frac{1}{4} 
\delta \lambda \mc^\varepsilon \, \varphi^4
\end{eqnarray}
with
\begin{equation}
\delta_\chi = Z_\chi - 1, \quad
\delta_\varphi = Z_\varphi - 1
\end{equation}
and 
\begin{eqnarray}
\label{dmu}
\delta \mu^2 &=& Z_\varphi\, \mu^2_B - \mu^2,
\\
\label{dy}
\delta y \mc^{\varepsilon/2} &=& Z_\chi \sqrt{Z_\varphi}\, y_B - y \mc^{\varepsilon/2},
\\
\label{dlambda}
\delta \lambda \mc^\varepsilon &=& Z_\varphi^2\, \lambda_B - \lambda \mc^\varepsilon.
\end{eqnarray}
\paragraph{Spontaneous symmetry breaking:} After renormalization, we 
implement spontaneous symmetry breaking by
assuming $\mu^2 < 0$ and splitting $\varphi$ into
\begin{equation}\label{split}
\varphi = v \mc^{-\varepsilon/2} + h,
\end{equation}
where $v$ is given by equation~(\ref{v}).
The factor $\mc^{-\varepsilon/2}$ ensures that $v$ has the
dimension of a mass. 
With equation~(\ref{split}), the full Lagrangian reads
\begin{subequations}
\begin{eqnarray}
\mathcal{L}_B &=& i \bar {\chi}_L \gamma^\mu \partial_\mu \chi_L +
\frac{1}{2} \left( \partial_\mu h \right) \left( \partial^\mu h \right) 
\\ && \label{yuk}
+\left( \frac{1}{2}\,y \mc^{\varepsilon/2}\, \chi_L^T C^{-1} \chi_L h +
\frac{1}{2}\, m_0\, \chi_L^T C^{-1} \chi_L + \mbox{H.c.} \right) 
\\ && \label{yuk-shift}
+\left( \frac{1}{2}\,\delta y \mc^{\varepsilon/2} \, \chi_L^T C^{-1} \chi_L h +
\frac{1}{2}\, \delta y v\, 
\chi_L^T C^{-1} \chi_L + \mbox{H.c.} \right) 
\\ &&
+ \, \delta_\chi i \bar {\chi}_L \gamma^\mu \partial_\mu \chi_L +
\delta_\varphi 
\frac{1}{2} \left( \partial_\mu h \right) \left( \partial^\mu h \right)
- V - \delta V
\end{eqnarray}
\end{subequations}
with
\begin{subequations}
\label{h^x}
\begin{eqnarray}
V + \delta V &=&
\left( - \frac{1}{4} \lambda v^4 + 
\frac{1}{2} \delta\mu^2 v^2 
+ \frac{1}{4} \delta\lambda\, v^4 
\right)  \mc^{-\varepsilon}
\\ && \label{1}
+ \left( \delta \mu^2 v + \delta \lambda\,
v^3 \right) \mc^{-\varepsilon/2} h 
\\ && \label{2}
+ \frac{1}{2} \left( M_0^2 + \delta\mu^2 + 3\,\delta \lambda\, v^2 \right) h^2
\\ && \label{3}
+ \left( \lambda v + \delta \lambda\, v
\right) \mc^{\varepsilon/2} h^3 
\\ && \label{4}
+ \frac{1}{4} \left( \lambda + \delta\lambda \right)
 \mc^\varepsilon h^4.
\end{eqnarray}
\end{subequations}
The terms of the scalar potential, including its counterterms, are listed in
equation~(\ref{h^x}) in ascending powers of $h$.
The appearance of a term linear in $h$ is analogous to that of the
linear $\sigma$-model~\cite{gervais}.
\paragraph{Renormalization conditions:} 
Here we state the 
five 
conditions for fixing the five parameters 
$\delta y$, $\delta \lambda$, $\delta \mu^2$, $\delta_\chi$ and 
$\delta_\varphi$ in the counter\-terms. 
\begin{enumerate}
\item
For simplicity, 
the $\overline{\mathrm{MS}}$ prescription is used to determine 
$\delta y$ and $\delta \lambda$, \textit{i.e.}\ the term proportional to
\begin{equation}
c_\infty = \frac{2}{\varepsilon} - \gamma + \ln(4\pi)
\end{equation}
is subtracted from the fermion vertex and the scalar four-point
function, respectively.
\item
The term linear in $h$, equation~(\ref{1}), induces a scalar VEV,
\textit{i.e.}\ a contribution to the scalar one-point function. We
choose $\delta\mu^2$ in such a way that 
the one-point function of the scalar field $h$
vanishes.
\item
The wave-function renormalization constants $\delta_\chi$ and
$\delta_\varphi$ are determined such that the 
residua of the fermion
and scalar propagators, respectively, are both one.
\end{enumerate}

As stressed in the introduction, we want to renormalize the VEV, so we
have to switch from $\delta \mu^2 $ to $\delta v$. 
Of course, this procedure makes only sense in the broken phase of the model.
Notice that in general we can define $\delta v$ via
\begin{equation}
v_B = \sqrt{-\frac{\mu^2_B}{\lambda_B}},
\quad
v = \sqrt{-\frac{\mu^2}{\lambda}}
\quad \mbox{and} \quad 
\delta v \mc^{-\varepsilon/2} = Z_\varphi^{-1/2} v_B - v
\mc^{-\varepsilon/2},
\end{equation}
in analogy to $\delta y$ and $\delta \lambda$. 
Starting from $v_B$ above, with $\mu^2_B$ and $\lambda_B$ from
equations~(\ref{dmu}) and~(\ref{dlambda}), respectively, 
we obtain
\begin{equation}\label{dva}
\delta v = 
v \sqrt{\frac{1+ \delta \mu^2/\mu^2}{1+\delta \lambda/\lambda}} - v,
\end{equation}
which allows us to trade $\delta \mu^2$ for $\delta v$.

At lowest order, equation~(\ref{dva}) yields
\begin{equation}\label{dv}
\delta v = - \frac{1}{M_0^2}
\left( \delta \mu^2 v + \delta \lambda\, v^3 \right). 
\end{equation}
Using this equation to replace $\delta \mu^2$ by $\delta v$, the
counterterms linear and quadratic in $h$, \textit{i.e.}\ 
equations~(\ref{1}) and~(\ref{2}), respectively, assume the form
\begin{equation}\label{sc}
-M_0^2\, \delta v \mc^{-\varepsilon/2}\, h + 
\frac{1}{2} \left(-2\lambda v\, \delta v + 
2 \delta \lambda\,v^2 \right) h^2.
\end{equation}
At one-loop order, there are two tadpole contributions
to the scalar one-point function, namely a fermionic contribution
$T_\chi$ and a scalar contribution $T_h$. Thus the associated 
renormalization condition is~\cite{fleischer,gervais,peskin}
\begin{equation}\label{T}
\raisebox{-6mm}{\includegraphics[width=0.42cm,angle=0]{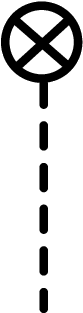}}
\;\;
+
\raisebox{-6mm}{\includegraphics[width=1.0cm,angle=0]{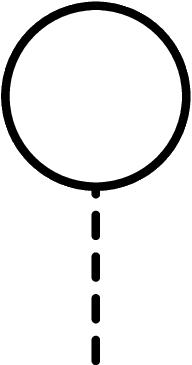}}
+
\raisebox{-6mm}{\includegraphics[width=1.0cm,angle=0]{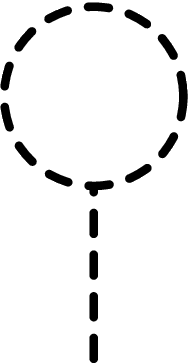}}
= \delta v \mc^{-\varepsilon/2} + T_\chi + T_h = 0,
\end{equation}
with the fermion and scalar contributions
given by
\begin{subequations}
\label{tadpoles}
\begin{eqnarray}
T_\chi & = & - \frac{2y\varsigma}{16\pi^2} 
\mc^{-\varepsilon/2}\, \frac{m_0^3}{M_0^2} 
\left( c_\infty + 1 - \ln \frac{m_0^2}{\mc^2} \right),
\\
T_h & = & \hphantom{-}\frac{3\lambda}{16\pi^2} \mc^{-\varepsilon/2}\, v
\left( c_\infty + 1 - \ln \frac{M_0^2}{\mc^2} \right),
\end{eqnarray}
\end{subequations}
respectively, where $\varsigma$ is defined as 
\begin{equation}\label{varsigma}
\varsigma = \left\{
\begin{array}{cl}
1 & \mbox{for a Majorana fermion loop,} \\
2 & \mbox{for a Dirac fermion loop.}
\end{array} \right.
\end{equation}
The first contribution to equation~(\ref{T}) stems from
the counterterm linear in $h$ in equation~(\ref{sc}).
\paragraph{Mass parameters:} It is useful to summarize the mass parameters
occurring in the model.
\begin{itemize}
\item
Mass scale of dimensional regularization: $\mc$.
\item
Mass squared of the scalar: $M^2 = M_0^2 + M_1^2$, where 
$M_0^2 = 2 \lambda v^2$ is the tree-level mass squared and $M_1^2$ the
one-loop correction.
\item
Mass of the fermion: $m = m_0 + m_1$, with the tree level mass $m_0 = yv$ and
its one-loop correction $m_1$.
\end{itemize}
\paragraph{Notes on Feynman rules for Majorana fermions:}
The Yukawa interaction term in equation~(\ref{yuk}) can be reformulated as
\begin{equation}\label{yy}
\frac{1}{2}\, \chi_L^T C^{-1} \chi_L (v \mc^{-\varepsilon/2}+ h) + \mbox{H.c.} = 
-\frac{1}{2}\, \bar\chi \chi ( v \mc^{-\varepsilon/2} + h )
\quad \mbox{with} \quad
\chi = \chi_L + (\chi_L)^c,
\end{equation}
with the charge conjugation indicated by the superscript.
For simplicity we have left out $y \mc^{\varepsilon/2}$.
Since in a Wick contraction a field $\chi$ can be contracted with both Majorana
fermion fields of the Yukawa interaction vertex, the
factor~$1/2$ in equation~(\ref{yy}) is always canceled when the
fermion line is \emph{not} closed. However,
with our convention for the Yukawa interaction, in every closed fermion loop
one factor $1/2$ remains~\cite{denner}. 

Note that for a Dirac fermion field $\chi$ we have defined
the Yukawa coupling \emph{without} the factor $1/2$.  
Therefore, with our convention there is no difference between the Majorana and
Dirac case for fermion lines which are not closed. 
However, a closed loop of a Dirac fermion
does \emph{not} have a factor $1/2$ in this convention. 
Therefore, in order to unify the treatment of Majorana and Dirac
fermions, we have introduced a factor $\varsigma$ defined in
equation~(\ref{varsigma}).

\section{Fermion selfenergy}
\label{fermion}
The terms contributing to the renormalized
fermion selfenergy at one-loop order are
\begin{equation}
\Sigma(p) = \Sigma_{\mbox{\scriptsize 1-loop}}(p) - \delta_\chi\, \slashed{p}
+ \delta_m^{(\chi)} + 
y \left\{ \delta v + \mc^{\varepsilon/2} \left( T_\chi + T_h \right) \right\}
\end{equation}
with
\begin{equation}\label{dm-chi}
\delta_m^{(\chi)} = v\,\delta y.
\end{equation}
The one-loop contribution is 
\begin{equation}
\Sigma_{\mbox{\scriptsize 1-loop}}(p) = \frac{y^2}{16\pi^2} \left[
-c_\infty \left( \frac{1}{2} \slashed{p} + m_0 \right) +
\int_0^1 \mathrm{d}x \left( x\slashed{p} + m_0 \right) 
\ln \frac{\Delta(p^2)}{\mc^2} \right]
\end{equation}
with
\begin{equation}
\Delta(p^2) = x M_0^2 + (1-x) m_0^2 -x(1-x)p^2.
\end{equation}
Due to the renormalization condition~(\ref{T}), the last term in
$\Sigma(p)$ vanishes.

The renormalized fermion selfenergy has the structure
\begin{equation}
\Sigma(p) = A(p^2) \slashed{p} + B(p^2) m.
\end{equation}
Both $A$ and $B$ are dimensionless and must be finite.
Obviously, $A$ can be made finite by an appropriate choice of
$\delta_\chi$, but for $B$ we do not have such a freedom because
\begin{equation}
\delta y = \frac{y^3}{16\pi^2} \,c_\infty
\end{equation}
is already determined by the renormalization of the fermion vertex. Therefore,
consistency requires that $\delta_m^{(\chi)}$ of equation~(\ref{dm-chi})
with this $\delta y$ makes $B$ finite, which is
indeed the case.

As emphasized before, 
in our renormalization scheme the masses are no free parameters but 
calculable in terms of the 
parameters of the model. Therefore, imposing the condition
of residue one at the pole of the fermion propagator~\cite{aoki},
\begin{subequations}
\begin{align}
\label{residue_one}
&A(m^2) + 2 m^2 \left[ A'(m^2) + B'(m^2) \right] = 0, 
\end{align}
automatically leads to a formula for the determination of the pole mass:
\begin{align}
\label{pole_mass}
&m = m_0 + m \left[ A(m^2) +  B(m^2) \right].
\end{align}
\end{subequations}
In equation~(\ref{residue_one}), the prime indicates the derivative with
respect to $p^2$.
Since we are satisfied with a one-loop computation, we can replace $m$ by
$m_0$ in equation~(\ref{residue_one}) and on the right-hand side of 
equation~(\ref{pole_mass}). 
Equation~(\ref{residue_one}) determines $\delta_\chi$ as
\begin{equation}
\delta_\chi = \frac{y^2}{16\pi^2} \left[ 
-\frac{1}{2} c_\infty + \int_0^1 \mathrm{d}x\, x
\ln \frac{\Delta(m_0^2)}{\mc^2}
-2m_0^2 \int_0^1 \mathrm{d}x\, \frac{x(1-x^2)}{\Delta(m_0^2)} \right].
\end{equation}
Finally, with this $\delta_\chi$ and using equation~(\ref{pole_mass}) we
obtain 
\begin{subequations}
\begin{equation}\label{final-m1}
m = m_0 + m_1 = 
m_0 \left[ 1 + \frac{y^2}{16\pi^2} \left(
2m_0^2 \int_0^1 \mathrm{d}x\, \frac{x(1-x^2)}{\Delta(m_0^2)} + 
\int_0^1 \mathrm{d}x \ln \frac{\Delta(m_0^2)}{\mc^2} \right) \right]
\end{equation}
with
\begin{equation}
\Delta(m_0^2) = x M_0^2 + (1-x)^2 m_0^2.
\end{equation}
\end{subequations}
In equation~(\ref{final-m1}) $m_1$ denotes the one-loop mass.

\section{Scalar selfenergy}
\label{scalar}
\begin{figure}[htb]
\centerline{%
\includegraphics[width=0.8\textwidth,angle=0]{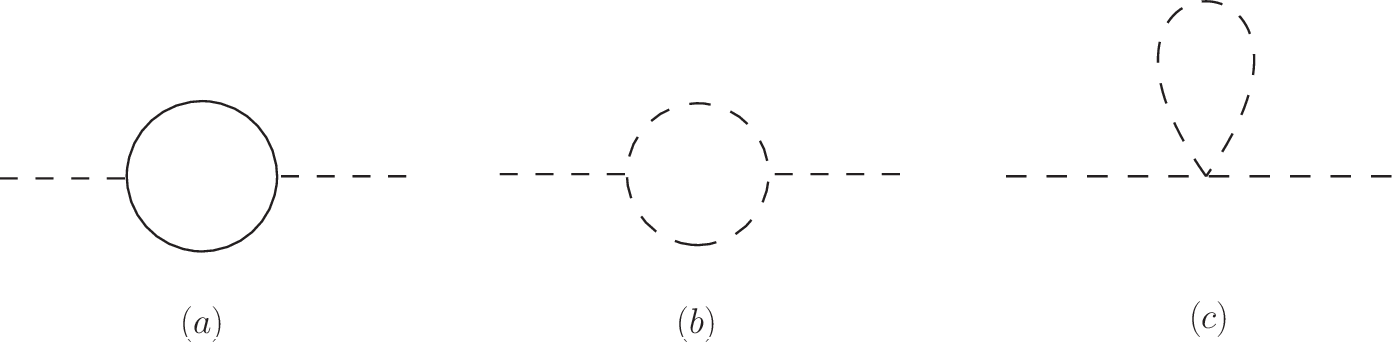}}
\caption{The Feynman diagrams of the three one-loop contributions
to the scalar selfenergy.}\label{scalar-selfenergy-graphs}
\end{figure}

The scalar selfenergy consists of the terms
\begin{equation}
\Pi(p^2) = \Pi_{\mbox{\scriptsize 1-loop}}(p^2) - p^2 \delta_\varphi +
\delta_m^{(h)} + 6\lambda v
\left\{ \delta v +  \mc^{\varepsilon/2} \left( T_\chi + T_h \right) \right\}.
\end{equation}
Just as for the fermionic selfenergy, the tadpole contributions are
canceled by $\delta v$. 
According to equation~(\ref{sc}), 
the quantity $\delta_m^{(h)}$ is given by
\begin{equation}\label{dm}
\delta_m^{(h)} = -2\lambda v\, \delta v + 
2 \delta \lambda\,v^2.
\end{equation}
The one-loop contribution has three terms,
\begin{equation}
\Pi_{\mbox{\scriptsize 1-loop}}(p^2) =
\Pi_{\mbox{\scriptsize 1-loop}}^{(a)}(p^2) +
\Pi_{\mbox{\scriptsize 1-loop}}^{(b)}(p^2) +
\Pi_{\mbox{\scriptsize 1-loop}}^{(c)}(p^2),
\end{equation}
referring to loops induced by the interaction terms of
equations~(\ref{yuk}),~(\ref{3}) and~(\ref{4}), respectively.
The corresponding Feynman diagrams are shown in
figure~\ref{scalar-selfenergy-graphs}. 
The results of the one-loop computation are
\begin{subequations}
\begin{eqnarray}
\Pi_{\mbox{\scriptsize 1-loop}}^{(a)}(p^2) &=&
\frac{2y^2\varsigma}{16\pi^2} \int_0^1 \mathrm{d}x\, \Delta_f(p^2) \left(
3 c_\infty + 1- 3 \ln \frac{\Delta_f(p^2)}{\mc^2} \right),
\label{a} \\
\Pi_{\mbox{\scriptsize 1-loop}}^{(b)}(p^2) &=&
-\frac{9 \lambda}{16\pi^2} M_0^2 \left( c_\infty - 
\int_0^1 \mathrm{d}x \ln \frac{\Delta_s(p^2)}{\mc^2} \right),
\label{b} \\
\Pi_{\mbox{\scriptsize 1-loop}}^{(c)}(p^2) &=&
-\frac{3\lambda}{16\pi^2} M_0^2 \left( c_\infty + 1 - \ln
\frac{M_0^2}{\mc^2} \right),
\label{c}
\end{eqnarray}
\end{subequations}
with
\begin{equation}
\Delta_f(p^2) = m_0^2 - x(1-x) p^2
\quad \mbox{and} \quad
\Delta_s(p^2) = M_0^2 - x(1-x) p^2.
\end{equation}

In $\delta_m^{(h)}$ the counterterm for the quartic scalar coupling is
required. It is obtained from the scalar four-point function as  
\begin{equation}\label{lambda-infty}
\delta \lambda = \frac{1}{16\pi^2} c_\infty \left( 9
\lambda^2 - 2 \varsigma y^4 \right).
\end{equation}

The renormalized scalar propagator has residue 
one at the physical mass
$M^2$ and the physical mass is determined by its pole. 
This means that the selfenergy has to fulfill
\begin{equation}
\Pi'(M^2) = 0 
\quad \mbox{and} \quad
M^2 = M_0^2 + \Pi(M^2),
\end{equation}
where the prime indicates the derivative with respect to $p^2$.
Since we perform a one-loop computation, we can replace $M^2$ in both 
$\Pi'(M^2)$ and $\Pi(M^2)$ by $M_0^2$. Thus we find
\begin{equation}
\delta_\varphi = \Pi'_{\mbox{\scriptsize 1-loop}}(M_0^2)
\quad \mbox{and} \quad
M^2 = M_0^2 + M_1^2 \quad \mbox{with} \quad
M_1^2 = \Pi(M_0^2).
\end{equation}
Explicitly, $\delta_\varphi$ is given by
\begin{equation}\label{delta-phi}
\delta_\varphi = 
\frac{2y^2\varsigma}{16\pi^2} \int_0^1 \mathrm{d}x\, x(1-x) \left(
-3 c_\infty + 2 + 3 \ln \frac{\Delta_f(M_0^2)}{\mc^2} \right) 
-\frac{9\lambda}{16\pi^2} \int_0^1 \mathrm{d}x\, \frac{x(1-x)}{1-x(1-x)}.
\end{equation}

On the one hand,
the finiteness in the derivative of $\Pi(p^2)$ has been enforced by
the choice of $\delta_\varphi$. Actually, only in
$\Pi_{\mbox{\scriptsize 1-loop}}^{(a)}(p^2)$ there is a divergent term
proportional to $p^2 c_\infty$ which is eliminated by a corresponding term in
$p^2 \delta_\varphi$, the other two contributions ($j=b,c$) to the scalar
selfenergy have no such term.
On the other hand, for the mass $M_1^2$ we have
no free counterterm anymore and this mass has to be finite via
$\delta_m^{(h)}$ which we have already determined earlier. 
Let us denote the $p^2$-independent terms proportional to $c_\infty$ in
$\Pi_{\mbox{\scriptsize 1-loop}}^{(j)}(p^2)$ 
by $\Pi^{(j)}_\infty$ ($j=a,b,c$).
These quantities  can be read off from equations~(\ref{a}), (\ref{b})
and~(\ref{c}), respectively: 
\begin{equation}
\Pi^{(a)}_\infty = \frac{6y^2\varsigma}{16\pi^2} m_0^2 c_\infty,
\quad
\Pi^{(b)}_\infty = -\frac{9\lambda}{16\pi^2} M_0^2 c_\infty,
\quad
\Pi^{(c)}_\infty = -\frac{3\lambda}{16\pi^2} M_0^2 c_\infty.
\end{equation}
The terms proportional to $c_\infty$ in $\delta_m^{(h)}$ are found by
considering equations~(\ref{tadpoles}) and~(\ref{lambda-infty}):
\begin{equation}
(\delta_m^{(h)})_\infty = \frac{1}{16\pi^2} \left[
-2 \lambda v \left( 2y\varsigma \frac{m_0^3}{M_0^2} - 3\lambda v \right) +
2 v^2 \left( 9 \lambda^2 - 2 y^4 \varsigma \right) \right] c_\infty.
\end{equation}
It is then easy to check that
\begin{equation}
\sum_{j=a,b,c} \Pi^{(j)}_\infty + (\delta_m^{(h)})_\infty = 0.
\end{equation}
We stress that, though due to the renormalization condition the tadpole
diagrams are canceled by $\delta v$ in the scalar selfenergy, they 
nevertheless play a crucial role in the cancellation of the infinities
in $M_1^2$ because
they occur in the counterterm $\delta_m^{(h)}$.

Collecting all terms, we obtain at one-loop order
\begin{eqnarray}
M^2 &=& M_0^2 \left\{
1 - \frac{y^2 \varsigma}{16 \pi^2} \left(
1 + 6 \frac{m_0^2}{M_0^2} \int_0^1 \mathrm{d}x\,
\ln \frac{\Delta_f(M_0^2)}{\mc^2} - 
2 \frac{m_0^2}{M_0^2} \ln \frac{m_0^2}{\mc^2} \right) \right.
\nonumber \\[1mm] && \left. \hphantom{xxxxx}
+ \frac{9 \lambda}{16 \pi^2} 
\left( \ln \frac{M_0^2}{\mc^2} - 3 + \frac{5\pi}{3\sqrt{3}}
\right) \right\}.
\label{final-M1}
\end{eqnarray}

\section{$\mc$-independence of the particle masses}
\label{independence}
The one-loop masses $m_1$ and $M_1^2$ depend explicitly on $\mc$. However,
the fermion and scalar masses, defined as pole masses, are  
physical observables and have to be independent of
$\mc$. Since we have performed a one-loop computation, 
independence means that the $\mc$-dependence must cancel at one-loop order which
implies that the implicit dependence of $m_0$ on $\mc$ must be
canceled by the explicit dependence of $m_1$ on $\mc$. The same has to hold
for $M_0^2$ and $M_1^2$. 
In the following we will demonstrate this and derive as well some useful other
relations concerning the $\mc$-dependence. 
The point of departure is the relationship between bare and renormalized
quantities and their counterterms~\cite{collins}:
\begin{subequations}\label{bb}
\begin{eqnarray}
y_B &=& \mc^{\varepsilon/2} \left( y + \delta y \right) Z_\chi^{-1}
Z_\varphi^{-1/2},
\label{yB} 
\\
\lambda_B &=& \mc^\varepsilon \left( \lambda + \delta \lambda \right) 
Z_\varphi^{-2},
\label{lambdaB}
\\
v_B &=& \mc^{-\varepsilon/2} \left( v + \delta v \right) 
Z_\varphi^{1/2}.
\label{vB}
\end{eqnarray}
\end{subequations}
Exploiting the fact that we do not go beyond one-loop order,
we obtain from equation~(\ref{bb}) formulae for the bare 
masses:\footnote{Note that, due to the appearance of $Z_\chi$ and $Z_\varphi$
  in these formulae, a gauge dependence will in general be introduced
  in the relation between the bare and renormalized masses,  
  as soon as gauge interactions are added to the
  Lagrangian---see for instance~\cite{fleischer}.} 
\begin{eqnarray}
&&
\hphantom{2}y_B v_B = \left( y + \delta y \right) 
\left( v + \delta v \right)
Z_\chi^{-1} = 
m_0 \left( 1 - \delta_\chi + \frac{\delta y}{y} + 
\frac{\delta v}{v} \right),
\\
&&
2\lambda_B v_B^2 = 2
\left( \lambda  + \delta \lambda \right)
\left( v + \delta v \right)^2 Z_\varphi^{-1} = 
M_0^2 \left( 1 - \delta_\varphi + \frac{\delta \lambda}{\lambda} + 
2\frac{\delta v}{v} \right).
\end{eqnarray}
Takin the derivative of the above equations with respect to $\mc$ and 
taking into account that bare quantities do not depend on $\mc$, we arrive at 
\begin{eqnarray}
0 &=& \mc \frac{\partial m_0}{\partial \mc} + m_0\, 
\mc \frac{\partial}{\partial \mc}
\left( - \delta_\chi + \frac{\delta y}{y} + 
\frac{\delta v}{v} \right),
\label{rg-m0} \\
0 &=& \mc \frac{\partial M_0^2}{\partial \mc} + M_0^2\, 
\mc \frac{\partial}{\partial \mc}
\left( -\delta_\varphi + \frac{\delta \lambda}{\lambda} + 
2\frac{\delta v}{v} \right).
\label{rg-M0}
\end{eqnarray}
To proceed further we bear in mind that $y^2$ and $\lambda$ correspond to
the same order in the loop expansion and that from equations~(\ref{yB})
and~(\ref{lambdaB}) we obtain at lowest order 
$\mc \frac{\partial y}{\partial \mc} = - \varepsilon y/2 + 
\mathcal{O}(y^3)$
and
$\mc \frac{\partial \lambda}{\partial \mc} = - \varepsilon \lambda +
\mathcal{O}(\lambda^2)$.
Then, with the results of sections~\ref{fermion} and~\ref{scalar}, 
it is straightforward to derive, at order $y^2$, 
\begin{equation}
\mc \frac{\partial}{\partial \mc} \delta_\chi =
\mc \frac{\partial}{\partial \mc} \delta_\varphi =
\mc \frac{\partial}{\partial \mc} \frac{\delta v}{v} = 0
\end{equation}
and
\begin{equation}
\mc \frac{\partial}{\partial \mc} \frac{\delta y}{y} = 
- \frac{2y^2}{16\pi^2},
\quad
\mc \frac{\partial}{\partial \mc} \frac{\delta \lambda}{\lambda} = 
\frac{1}{16\pi^2} \left( -18 \lambda + 4\varsigma\, \frac{y^4}{\lambda} \right).
\end{equation}
Plugging these results into equations~(\ref{rg-m0}) and~(\ref{rg-M0}), we find
the \emph{implicit} dependence of the tree-level masses on $\mc$: 
\begin{subequations}
\label{rg1}
\begin{eqnarray}
\mc \frac{\partial m_0}{\partial \mc}
&=& \frac{2y^2}{16\pi^2} \, m_0,
\label{rg1-m0}
\\
\mc \frac{\partial M_0^2}{\partial \mc} 
&=& 
\frac{1}{16\pi^2} \left( 18 \lambda M_0^2 - 8\varsigma y^2 m_0^2 \right).
\label{rg1-M0}
\end{eqnarray}
\end{subequations}
The latter expression is, of course, proportional to $\delta \lambda$
of equation~(\ref{lambda-infty}).
With equation~(\ref{rg1}), it 
is now trivial to see that 
the derivative of
$m_1$, equation~(\ref{final-m1}), and of
$M_1^2$, equation~(\ref{final-M1}), 
with respect to the explicit dependence on $\mc$ 
leads to expressions with signs opposite to those of 
equations~(\ref{rg1-m0}) and~(\ref{rg1-M0}), respectively.
Therefore, $m$ and $M^2$ are independent of $\mc$ 
at order $y^2$ and $\lambda$, which proves our statement in the beginning of
this section.

Finally, for completeness we also present the beta functions of $y$,
$\lambda$~\cite{cheng} and~$v$ at one-loop order:
\begin{equation}
\mc \frac{\partial y}{\partial \mc} = 
-\frac{\varepsilon}{2}\,y + 
\frac{2y^3}{16\pi^2},
\quad
\mc \frac{\partial \lambda}{\partial \mc} = 
-\varepsilon \lambda + 
\frac{1}{16\pi^2} \left( 18 \lambda^2 - 4\varsigma y^4 \right),
\quad
\mc \frac{\partial v}{\partial \mc} 
= \frac{\varepsilon}{2}\,v.
\end{equation}

\section{Conclusions}
\label{concl}
In this paper we have discussed renormalization in a minimalist model with 
a spontaneously broken discrete symmetry. 
We have employed a renormalization scheme for the broken phase
which renormalizes the VEV such that the one-point function of the shifted
field $h$ defined in equation~(\ref{split}) vanishes; 
for simplicity we have renormalized the Yukawa coupling constant $y$ and the
quartic scalar coupling constant $\lambda$ by the $\overline{\mbox{MS}}$ scheme.
The renormalization condition for the VEV implies that tadpole contributions
to the selfenergies are canceled by the renormalization parameter 
$\delta v$ of the VEV.
In this model, since there is no fermion mass term in the unbroken Lagrangian,
it is obvious that the one-loop fermion mass gets renormalized by the
renormalization parameter $\delta y$. Concerning the scalar mass, 
the mass counterterm of the scalar contains explicitly the tadpole
contributions in $\delta_m^{(h)}$---see equation~(\ref{dm}), therefore, both
$\delta v$ and $\delta \lambda$ are needed for the finiteness of the scalar
mass $M^2$. Moreover, $M^2$ also receives a finite tadpole 
contribution.\footnote{Such finite contributions of tadpoles to fermion masses
  have recently been discussed in the context of the Standard Model---see for
  instance~\cite{jegerlehner}.} 

We have considered this toy model, which does neither have
the complications due to gauge interactions~\cite{sperling} nor those due to
propagator mixing~\cite{kniehl}, in view of a future
application to renormalizable flavour models.
We emphasize that our renormalization scheme has the virtue of 
facilitating a clear formulation of the hierarchy or fine-tuning problem 
because it allows the comparison of the tree-level masses with their radiative
corrections without obfuscation by a cut-off. In addition, 
it seems straightforward to extend the model to the case of an arbitrary
number of fermion and scalar fields.
Finally, we mention that there is an interesting difference 
between the Majorana and Dirac fermion contribution to the one-loop mass of 
equation~(\ref{final-M1}) due to occurrence of  
$\varsigma$---see equation~(\ref{varsigma}) for its definition.

\vspace{3mm}

\noindent
\textbf{Acknowledgments:}
This work is supported by the Austrian
Science Fund (FWF), Project No.\ P~24161-N16. W.G. thanks 
Christoph Bobeth and Martin Gorbahn for discussions on the
VEV renormalization during the \textit{Workshop ``Towards the
  Construction of the Fundamental Theory of Flavour''} at the
Institute for Advanced Study TUM in Munich.

\end{document}